\documentstyle[12pt,aaspp4]{article}
\begin{document}
\newcommand{\up}[1]{\ifmmode^{\rm #1}\else$^{\rm #1}$\fi}
\newcommand{\zdot}{\makebox[0pt][l]{.}}
\newcommand{\upd}{\up{d}}
\newcommand{\uph}{\up{h}}
\newcommand{\upm}{\up{m}}
\newcommand{\ups}{\up{s}}
\newcommand{\arcd}{\ifmmode^{\circ}\else$^{\circ}$\fi}
\newcommand{\arcm}{\ifmmode{'}\else$'$\fi}
\newcommand{\arcs}{\ifmmode{''}\else$''$\fi}

\title{The Araucaria Project. A Distance Determination to the Local Group Spiral M33 from Near-Infrared 
Photometry of Cepheid Variables
\footnote{Based on observations obtained with the ESO VLT for 
Programme 382.D-0469(A)}
}
\author{Wolfgang Gieren}
\affil{Universidad de Concepci{\'o}n, Departamento de Astronomia,
Casilla 160-C, Concepci{\'o}n, Chile}
\authoremail{wgieren@astro-udec.cl}
\author{Marek G{\'o}rski}  
\affil{Warsaw University Observatory, Al. Ujazdowskie 4, 00-478, Warsaw,
Poland}
\authoremail{mgorski@astrouw.edu.pl}
\author{Grzegorz Pietrzy{\'n}ski}
\affil{Universidad de Concepci{\'o}n, Departamento de Astronomia,
Casilla 160-C, Concepci{\'o}n, Chile}
\affil{Warsaw University Observatory, Al. Ujazdowskie 4, 00-478, Warsaw,
Poland}
\authoremail{pietrzyn@hubble.cfm.udec.cl}

\author{Piotr Konorski}
\affil{Warsaw University Observatory, Al. Ujazdowskie 4, 00-478, Warsaw,
Poland}
\authoremail{piokon@astrouw.edu.pl}

\author{Ksenia Suchomska}  
\affil{Warsaw University Observatory, Al. Ujazdowskie 4, 00-478, Warsaw,
Poland}
\authoremail{piokon@astrouw.edu.pl}

\author{Dariusz Graczyk}   
\affil{Universidad de Concepci{\'o}n, Departamento de Astronomia,
Casilla 160-C, Concepci{\'o}n, Chile}
\authoremail{darek@astro-udec.cl}

\author{Bogumil Pilecki}   
\affil{Universidad de Concepci{\'o}n, Departamento de Astronomia,
Casilla 160-C,Concepci{\'o}n, Chile}
\affil{Warsaw University Observatory, Al. Ujazdowskie 4, 00-478, Warsaw,
Poland}
\authoremail{pilecki@astrouw.edu.pl}

\author{Fabio Bresolin}
\affil{Institute for Astronomy, University of Hawaii at Manoa, 2680 Woodlawn 
Drive, 
Honolulu HI 96822, USA}
\authoremail{bresolin@ifa.hawaii.edu}
\author{Rolf-Peter Kudritzki}
\affil{Institute for Astronomy, University of Hawaii at Manoa, 2680 Woodlawn 
Drive,
Honolulu HI 96822, USA}
\authoremail{kud@ifa.hawaii.edu}
\author{Jesper Storm}
\affil{Leibniz-Institut fuer Astrophysik Potsdam (AIP), An der Sternwarte 16, D-14482
Potsdam, Germany}
\authoremail{jstorm@aip.de}
\author{Paulina Karczmarek}
\affil{Warsaw University Observatory, Al. Ujazdowskie 4, 00-478, Warsaw,  
Poland}
\authoremail{pkarczma@astrouw.edu.pl}
\author{Alex Gallenne}
\affil{Universidad de Concepci{\'o}n, Departamento de Astronomia, Casilla 160-C,
Concepci{\'o}n, Chile}
\authoremail{dgallenne@astro-udec.cl}
\author{Paula Calder{\'o}n}
\affil{Universidad de Concepci{\'o}n, Departamento de Astronomia, Casilla 160-C,
Concepci{\'o}n, Chile}
\authoremail{pcalderon@astro-udec.cl}
\author{Doug Geisler}
\affil{Universidad de Concepci{\'o}n, Departamento de Astronomia, Casilla 160-C,
Concepci{\'o}n, Chile}
\authoremail{dgeisler@astro-udec.cl}

\begin{abstract}
Motivated by an amazing range of reported distances to the nearby Local Group spiral galaxy M33,
we have obtained deep near-infrared photometry for 26 long-period Cepheids in this galaxy
with the ESO VLT. From the data we constructed period-luminosity
relations in the J and K bands which together with previous optical VI photometry for the
Cepheids by Macri et al. were used to determine the true distance modulus of M33, and the
mean reddening affecting the Cepheid sample with the multiwavelength fit method developed
in the Araucaria Project. We find a true distance modulus of 24.62 for M33, with a total 
uncertainty of $\pm$ 0.07 mag which is dominated by the uncertainty on the photometric
zero points in our photometry. The reddening is determined as E(B-V)=0.19 $\pm$ 0.02, in agreement 
with the value used by the HST Key Project of Freedman et al. but in some discrepancy with other
recent determinations based on blue supergiant spectroscopy and an O-type eclipsing binary
which yielded lower reddening values. Our derived M33 distance modulus is extremely insensitive
to the adopted reddening law. We show that the possible effects of metallicity and
crowding on our present distance determination are both at the 1-2\% level and therefore minor
contributors to the total uncertainty of our distance result for M33.
\end{abstract}

\keywords{distance scale - galaxies: distances and redshifts - galaxies:
individual(M33)  - stars: Cepheids - infrared photometry}

\section{Introduction} 
As the second-nearest spiral galaxy, the Triangulum Galaxy M33 is one of the primary calibrators
for secondary distance indicators including the Tully-Fisher relation. Therefore an accurate
determination of its distance is a crucial step in the process of building the cosmic distance ladder.
Due to its relative proximity, basically all stellar methods of distance determination can be,
and have been employed to measure the distance to M33. The galaxy is therefore a perfect
laboratory to compare the distances derived from different techniques, and this way discover
systematic uncertainties affecting them.

While Cepheid variables were discovered in M33 as early as in the 1920s (Hubble 1926), it was
only 60 years later that reasonable estimates of the M33 distance from Cepheids
became available. Madore et al. (1985) were the first to employ random-phase near-infrared (H-band) 
photometry of 15 Cepheids to measure a M33 distance modulus of 24.5 $\pm$ 0.2 mag, or 24.3 mag 
with a reddening correction adopted from blue-band data of Sandage (1983). Freedman et al. (1991)
determined the distance to M33 using ground-based BVRI CCD photometry of 10 bright Cepheids, obtaining
an absorption-corrected true distance modulus of 24.64 $\pm$ 0.09 mag. Later, this value was revised
to 24.56 $\pm$ 0.10 adopting a mean reddening of E(B-V)=0.20 for the Cepheids (Freedman
et al. 2001). Lee et al. (2002) used single-epoch I-band observations of 21 Cepheids in M33 obtained with
HST/WFPC2 to determine the M33 true distance modulus as 24.53 $\pm$0.14 (random) $\pm$0.13 (systematic),
adopting the same mean reddening used by Freedman et al. (2001). All these values for the distance of M33
were tied to an adopted true distance modulus of 18.50 for the LMC. 

While the number of Cepheids known in M33 was very small prior to 2001, it was the DIRECT Project
(Macri et al. 2001) which increased this number very significantly to 251 variables, presenting BVI light 
curves for this sample
obtained at the Fred Lawrence Whipple Observatory (FLWO) 1.2 m telescope. Given the small aperture 
and limited spatial resolution of this
telescope, the M33 Cepheid light curves of Macri et al. are naturally quite noisy. 
However, the periods of the
variables could be determined quite accurately from these data. Lee et al. (2002) were able to take
advantage of these periods for their sample of DIRECT Cepheids which they observed with HST.

In the past 12 years, a number of new distance determinations to M33 from a variety of methods have
become available which we summarize for convenience in Table 1. They span a surprisingly large 
interval of 24.3-24.9 in the true distance modulus, with a 30\% difference between the extreme values
which hints at important systematic uncertainties inherent in several of these determinations. The
true distance modulus of M33, the second-nearest spiral galaxy and located in the Local Group,
is therefore presently clearly an ill-determined number, which is a cause of serious concern. 
The motivation for the present study was to alleviate this problem by performing for the first time
deep near-infrared imaging of Cepheids in the J and K bands,
and applying the multiwavelength optical-NIR method developed in the Araucaria Project (Gieren et al. 2005a,
2005b; Pietrzynski et al. 2006) to derive an accurate distance and reddening for M33.
 This is very important because different
assumptions about reddening are a prime source for the discrepancies among the various distance
determinations for M33 listed in Table 1.

As the first modern near-infrared Cepheid study in M33 since the pioneering
work of Madore et al. some 30 years ago, we consider this work as long overdue and helpful to understand
the discrepancies mentioned above, and improve the status of M33 as an anchor point in the extragalactic
distance scale.

\section{Observations, Data Reduction and Calibration}
The observations were collected with the 8.2 m ESO Very Large Telescope equipped 
with the HAWK-I wide-field infrared camera (Kissler-Patig et al. 2008).
The field of view of the camera is 7.5 $\times$ 7.5 arcmin with a pixel scale of 0.106". 
Based on the catalog of Cepheids in M33 obtained by the DIRECT project 
(Macri et al. 2001) one HAWK-I field, containing a relatively large sample of Cepheids
spanning a large range of pulsational periods, and located sufficiently far away from the 
very dense central regions of M33 was selected (see Fig. 1). This field 
was observed once in J and six times in K band filters, respectively during the period 
between December 4 2008 and Dec 5 2009.  The final images were obtained from a co-addition 
of 12 and 6 dithered observations for J and K band, respectively. Each K band observation
consists of six 10 seconds expositions. In  the case of the J band observations 
two 30 s exposures were made at each telescope position. Therefore the total exposure 
time for the J and K band observations were 12 and 6 minutes, respectively.
In order to accurately subtract the sky a relatively empty field located some 8 arcmin 
away from the selected Cepheid field was monitored between science exposures. 
The PSF photometry and calibrations were obtained in the same manner as in 
Pietrzynski et al. (2002). 

In order to accurately calibrate our observations onto the standard system 
we observed the same field in M33 with NTT+SOFI during two photometric nights 
together with a large number (12-16) of UKIRT standard stars (Hawarden et al. 2001) 
spanning a large range of colors. The calibration and reductions of the SOFI images were done 
in an exactly the same manner as in our previous work on infrared photometry of Cepheids in nearby 
galaxies (e.g. Pietyrzynski et al. 2006). The accuracy of the zero points in J and K bands
for each night, were estimated to be 0.06 and 0.03 mag respectively. The calibrated photometry of 
our shallow SOFI images taken during both nights agree each other within the estimated error 
of the corresponding zero point. 

Finally the HAWKI photometry was calibrated based on the commom stars identified 
on a given HAWKI chip and the SOFI image. In all cases the accuracy of this procedure 
was better than 0.02 mag. 

\section{The Cepheid Period-Luminosity Relations and the Distance to M33}
In Table 2, we present the journal of calibrated single-epoch J and multi-epoch K magnitudes of 26 long-period
classical Cepheids (periods between 6-74 days) we were able to identify
in our observed VLT/HAWKI field. The Cepheid identifications and the pulsation periods were adopted
from Macri et al. (2001).  In principle it would be possible to use these data together with
the V- and I-band light curves of the variables given by Macri et al. (2001), and the known periods
to apply the procedure outlined in Soszynski et al. (2005) to calculate the mean magnitudes of the
Cepheids in the J and K bands. However, the noisy optical light curves of the stars, the relatively low
precision of their pulsation periods and the long epoch difference between Macri's and our observations
would lead to large errors on the mean magnitudes, in the present case. 
Since the light curve amplitudes of classical Cepheids
in the near-infrared domain, however, are substantially reduced as compared to optical bands, random-phase 
J band observations and the mean from 3-6 K band observations  still do allow for a rather precise determination 
of Cepheid IR PL relations. The adopted mean magitudes used for P-L relations are listed in Table 3. 

In Figure 2, we show the J- and K-band PL relations based on the data in Table 2. It is appreciated
that the Cepheids rather homogeneously cover the period range between 6 and 74 days, yielding an excellent
period baseline for fitting PL relations to the data. Weighted least-squares fits to the data in Figure 2
yield slopes of the PL relations of -2.94 in J and -3.17 in K, with 1$\sigma$ uncertainties of 0.14 and
0.13, respectively. These values have to be compared with the slopes of the PL relations in J and K
determined by Persson et al. (2004) from a large sample of LMC Cepheids, which are -3.153 and -3.261,
respectively. Within the uncertainties, the PL relation slopes determined from our M33 Cepheid sample 
are consistent with the slope of the corresponding LMC Cepheid PL relation, once again confirming our
previous conclusion in the Araucaria Project that the slopes of the near-IR Cepheid PL relations in J,
and particularly in K are universal and not dependent on metallicity (this conclusion was 
independently confirmed by Storm et al. (2011a,b) from a comparison of Cepheid absolute magnitudes
in the Galaxy, LMC and SMC as determined from the Infrared Surface Brightness Technique). We therefore
adopt the approach used in the previous papers of our project to fit the Persson et al. LMC Cepheid
slopes to the data. This yields the following results:

J = -3.153 log P + (22.520 $\pm$ 0.045)  \\

K = -3.261 log P + (22.221 $\pm$ 0.036)    \\

Using the V- and I-band light curves of our current sample of 26 M33 Cepheids given by Macri et al. (2001), 
free fits
to the mean magnitudes versus log P diagrams yield slopes of -2.36 and -2.93 in V and I, respectively,
with 1$\sigma$ uncertainties of 0.21 and 0.15. The corresponding slopes for LMC Cepheids as determined
by the OGLE Project (Udalski et al. 1998) are -2.775 in V and -2.977 in I. There is excellent agreement
of the slopes in the I band whereas the deviation in V is at the 2$\sigma$ level, which seems tolerable.
Fitting the OGLE LMC Cepheid slopes to the data of Macri et al. yields the following equations:

V = -2.775 log P + (23.764 $\pm$ 0.066)   \\

I = -2.977 log P + (23.153 $\pm$ 0.046)    \\

Adopting a LMC distance modulus of 18.50, as in our previous papers, and as very recently confirmed
by our work on LMC eclipsing binaries (Pietrzynski et al. 2013), the zero points in the
previous equations translate into distance moduli of M33 of 24.70 in K, 24.73 in J, 25.06 in I
and 25.20 in V. As in the previous papers in this series, we adopt the extinction law of Schlegel
et al. (1998) and fit a straight line to the relation: 

$(m-M)_{0}$ = $(m-M)_{\lambda}$ - ${\rm A}_{\lambda}$ = $(m-M)_{\lambda}$ - ${\rm E}_{\rm B-V}$ ${\rm R}$

Using the reddened distance moduli derived above in the VIJK photometric bands and the R values
from Schlegel et al. (3.24, 1.96, 0.902, 0.367 in V,I, J and K, respectively), we obtain the following
results for the mean reddening affecting our Cepheid sample in M33, and the extinction-corrected distance modulus
of the galaxy:

${\rm E}_{\rm B-V}$ = 0.19 $\pm$ 0.02   \\

$(m-M)_{0}$ = 24.62 $\pm$ 0.03   \\

Using, instead of the Macri et al. (2001) data, the more recent M33 Cepheid VI data of Pellerin \& Macri (2011)
together with our JK data, the extinction-corrected distance modulus becomes 24.61 $\pm$ 0.03, and the reddening
0.20 $\pm$ 0.02. These values nicely agree with the above result. If we include in the analysis also the B data of Pellerin
\& Macri, the reddening value changes slightly to 0.18 $\pm$ 0.01, and the extinction-corrected distance modulus
becomes 24.62 $\pm$ 0.03. The conclusion is that we get, within a fraction of the statistical 1 $\sigma$ error,
identical results for the M33 extinction-corrected distance modulus, no matter if we use the older Macri et al. (2001),
or the more recent and abundant Pellerin \& Macri (2011) data in conjunction with our JK data presented in this paper.
This just reflects the fact that the near-infrared photometry strongly dominates our distance determination to M33.

Fig. 3 demonstrates that both the reddening, and the extinction-corrected distance modulus are well determined by the
fit to a straight line to the data. In Table 4, we summarize the results of this section, and we
present the extinction-corrected distance moduli calculated for each band with the reddening determined in this section.
The agreement with our adopted extinction-corrected distance modulus of 24.62 mag from the multiwavelength fit
is very good in each band. 

\section{Discussion}
The Cepheid distance to M33 derived in this paper is potentially affected by a number of systematic
uncertainties, in addition to the uncertainty on the photometric zero points discussed in section 2,
 which will now be discussed in turn. 

\subsection{Sample Selection}
We have chosen the observed VLT/HAWKI field in M33
with the criteria to maximize the number of long-period Cepheids contained in the field and
optimize their period distribution. We also chose the field as to be located at a medium
galactocentric distance where crowding of the Cepheids and its expected systematic effect
on the Cepheid photometry is reasonably small whereas at the same time the average expected
metallicity of the Cepheids is close to the metallicity of LMC Cepheids, which constitute
our adopted fiducial period-luminosity relations, in order to minimize the effect of metallicity
on the derived distance to M33. As can be seen in Fig. 2, the Cepheid sample in M33 used in
this study is quite large (26 variables), and it covers the broad period range from 6-74 days 
very homogeneously. The location of the Cepheids in the K vs. J-K color-magnitude diagram
shown in Fig. 4 confirms their nature as classical Cepheids. 

The observed dispersion on the PL diagram in the J band is likely dominated by the random-phase
nature of the observations in this band, where in a few cases Cepheids might be up to 0.25 mag brighter
or fainter than their mean magnitudes if our observation happened to occure close to maximum or
minimum light for these variables (the total light amplitude in J and K of a typical long-period Cepheid
is about 0.5 mag; see Persson et al. 2004). Since in K we have 2-5 observations at different epochs
for the variables, the data points in the K-band PL diagram should exhibit smaller scatter
around the ridge line, which is actually observed in Fig. 2.  

The contribution of the random photometric
errors of the magnitudes (see Table 2) on the scatter in Fig. 2 are negligible 
whereas unusually strong
dust absorption could be important in a few cases, particularly
in the J-band PL relation where the effect of dust absorption is stronger than in K. Our sample,
consisting of Cepheids all having periods longer than 6 days, should not contain any overtone
pulsators which would introduce additional scatter at the short-period end of the PL diagram
in Fig. 2 and a bias in the distance determination. Numerous studies (e.g. Soszynski et al. 2008)
have shown that at periods longer than about 5 days all classical Cepheids are fundamental mode
pulsators. As a test on this hypothesis we have done fits to the PL relations in Fig. 2 excluding
the Cepheids with period shorter than 10 days. This led to negligible changes in the zero points
of the relations, and in the distance to M33 providing strong evidence that our sample does indeed
not contain any overtone Cepheids.

\subsection{Reddening}
Without doubt reddening is one of the most serious sources of systematic error on distances to
galaxies determined with classical Cepheids or other young stars, and frequently the discrepancies 
between reported distance results for a given galaxy can be attributed to differences in the adopted 
reddenings. In the case of M33, all studies in Table 1 published later than 2001
 which have used optical photometry of
Cepheid variables to determine the distance have adopted a mean reddening value of 0.20 which was
derived in the HST Key Project final paper of Freedman et al. (2001). The resulting M33 true distance moduli 
from Cepheids cluster around 24.6. Interestingly, in their original study of M33 Cepheids
Freedman et al. (1991) had determined E(B-V)=0.10 and a true distance modulus of 24.64;
with the revised reddening of 0.20 adopted in the Key Project, they
would have obtained a shorter M33 true distance modulus of 24.32. The Galactic foreground reddening to M33
from the Schlegel et al. (1998) maps is E(B-V)=0.04 mag.

Contrasting with the 0.20 mag reddening value, two other recent studies employing young stars for a
distance determination to M33 have
found lower reddenings of E(B-V)=0.09 for an O-type eclipsing binary (Bonanos et al. 2006), and 
0.08 from a sample of 22 blue supergiant stars for which individual reddenings were determined 
spectroscopically from fits to their observed spectral energy distributions
(U et al. 2009), leading in both cases to distance moduli about 0.3 mag larger than the value 
determined from Cepheids. In the U et al. paper, a range of E(B-V) values between 0.02 and 0.16 was measured 
for the different
supergiants, demonstrating a considerable local variation of dust absorption in M33. This is also seen
in the E(B-V) values determined for H II regions in M33 which for a few objects exceed 0.6 mag,
although the mean value is close to 0.11 mag (Rosolowsky \& Simon 2008). From the data in that paper, 
H II regions with reddenings close to 0.20 are no exception. We also note here that Bresolin (2011) found
that for eleven H II regions in common with Rosolowsky and Simon, the mean E(B-V) values were 0.13 mag
larger, so that a value of E(B-V) in the range 0.20-0.25 for H II regions is certainly consistent with
modern data. In fact, Bresolin (2011) obtains a mean E(B-V) of 0.25 $\pm$ 0.17 from 25 H II regions studied
in his paper, suggesting that one can certainly not discard the possibility that the average E(B-V) value
corresponding to H II regions in M33 is $\sim$ 0.2.

The existence of strong local variations of the dust extinction intrinsic to M33 obviously
complicates the distance analysis from Population I objects and underlines
the crucial importance of deriving the reddening for a given sample of stars
with the utmost care and precision. In our Araucaria Project, we have shown in a number of previous 
papers that our adopted approach to use combined near-infrared and optical photometry, creating a very long
wavelength baseline, is capable of yielding a very robust determination of
the mean reddening affecting a given Cepheid sample
(Gieren et al. 2005b; Pietrzynski et al. 2006; Soszynski et al. 2006; Gieren et al. 2006; 
Gieren et al. 2008a,b; Gieren et al. 2009). The mean reddening of 0.19 $\pm$ 0.02 mag we have obtained
in this study for our Cepheid sample
agrees very well with the Freedman et al. (2001) determination from Cepheids but
is clearly more accurate due to the inclusion of near-infrared photometry in our analysis. 

In order to check the stability of our result for the extinction-corrected distance modulus on the adopted reddening law,
we have calculated the extinction-corrected distance and reddening as in Fig. 3, adopting different values for ${\rm R}_{V}$
ranging from 2.5 to 4.5, taking into account that local variations of the reddening law of such size
are not uncommom in star forming regions in spiral galaxies. The result is that the extinction-corrected distance modulus
is {\it extremely insensitive} to the adopted reddening law - it varies from 24.63 for ${\rm R}_{V}$=2.5 to
24.60 for a rather extreme value of ${\rm R}_{V}$=4.5, or just by 1.5\%, which is within the statistical
error of our adopted distance modulus of 24.62. The reddenings vary more strongly, from 0.242 $\pm$ 0.029
to 0.134 $\pm$ 0.016, for the different R values. Therefore an assumed high mean value for R can alleviate to some
extent the discrepancy of the photometric with the low spectroscopic reddenings of the blue supergiants in M33 analyzed by
U et al. (2009), but the discrepancy cannot be fully explained this way. The important conclusion of this
section however is the strong insensitivity of our distance result for M33 to the assumed reddening and reddening law,
which is mainly a consequence of having K-band photometry at our disposal, and underlines the importance
of including K-band photometry in the distance determinations with Cepheid variables.

\subsection{Metallicity Effects}
The effect of metallicity on Cepheid absolute magnitudes and the Period-Luminosity relation is still
poorly understood, but on the other hand obviously of crucial importance for a distance determination
based on Cepheid variables. While most researchers will agree that the slope of the PL relation
is metallicity independent, particularly in near- and mid-infrared photometric bands (e.g. Storm
et al. 2011a,b; Freedman et al. 2009), the effect of metallicity
on the zero point of the PL relations in different photometric bands has been widely disputed
in the past, with no consensus reached yet. The "classical" method to measure the metallicity effect 
has been to determine,
for a given spiral galaxy, the distance to a sample of Cepheids located close to the center of the galaxy 
(inner Cepheids),
and to another sample located in the outskirts of the same galaxy (outer Cepheids). The zero point difference
between the two samples is interpreted as due to the difference in the mean metallicities of the two
samples, the inner Cepheids being more metal-rich than the outer Cepheids. This way, it has generally been 
found that metal-rich Cepheids appear to be brighter, at the same period, than their more metal-poor
counterparts, however with the reported size of the effect varying enormously from a mild effect of about
-0.2 mag/dex (e.g. Kennicutt et al. 1998) up to a very strong effect close to -0.6 mag/dex (Gerke et al. 2011)
in the reddening-free V-I Wesenheit magnitude.
A fundamental problem in this approach is the choice of the metallicity gradient to be adopted
for this determination of the metallicity effect (eg. Kudritzki et al. 2012; Bresolin
et al. 2009) which strongly depends on the adopted calibration of H II region oxygen abundances.
Arguments have been given which seem to exclude a strong metallicity effect on Cepheid magnitudes
(Majaess et al. 2011) which are supported by the results for Cepheid distances to Milky Way and
Magellanic Cloud Cepheids from the infrared surface brightness technique (Storm et al. 2011a).

On the other hand, Romaniello et al. (2008) have found even a different {\it sign} of the metallicity effect
from high-resolution spectroscopic abundance determinations for Cepheids in the Milky Way and
Magellanic Clouds, casting doubts on the validity of the results produced by the "classical" technique,
which might be strongly affected by the crowding and blending affecting Cepheids in the central regions
of spiral galaxies making these Cepheids spuriously brighter than Cepheids located further away
from the center of the galaxy, through  significant photometric
contamination by unresolved companion stars (e.g. Majaess et al. 2012). In a recent paper, Freedman \& Madore (2011)
have related PL relation magnitude residuals for Magellanic Cloud Cepheids to their spectroscopic metallicities
as determined by Romaniello et al., for photometric bands ranging from U through the 8.0 micron Spitzer band,
yielding evidence that the metallicity effect varies sytematically with wavelength in size and sign. According
to their study, the metallicity effect on Cepheid absolute magnitudes is smallest, and consistent with zero,
in the near-infrared JHK bands, and particularly in K. This study has the advantage that the Cepheid magnitude
residuals can be assumed to be unbiased by blending and crowding, due to the relative proximity of the Clouds.

Even at HST resolution, galaxies
located further away than a few Megaparsecs are very difficult to resolve in their central regions which
makes photometric contamination as a significant contributor to the observed brightening of inner
Cepheids, particularly in massive spiral galaxies, very likely, and a separation of the effects belonging
to metallicity and crowding very difficult. In the particular case of M33, the relatively large magnitude
difference between inner and outer field Cepheids found by Scowcroft et al. (2009) is almost certainly
due to the low spatial resolution of their images and the resulting severe problems with crowding and
blending effects discussed in the next section, and is not caused by a significant difference in
the mean metallicities of these two samples. In our much higher-resolution HAWKI images we do not see
any evidence for a significant radial variation of the mean brightness of Cepheids of similar periods,
although admittedly the coverage of our HAWKI image in galactocentric distance in M33, and the number of Cepheids 
in the field is not large enough to reach very strong conclusions in this regard.

In order to handle the metallicity problem in the most practical way in the present study, we chose
our target field  as to reach an optimum compromise between avoiding crowding as much as possible, and
having a mean metallicity of the Cepheids in the field close to the one of LMC Cepheids
(-0.35 dex; Luck et al. 1998, Romaniello et al. 2008). Indeed, using the metallicities and
metallicity gradient in M33 as determined from blue
supergiants and H II regions  (U et al. 2009; their Figure 14), including metallicities 
for beat Cepheids determined by Beaulieu et al. (2006), the expected metallicity of Cepheids
at the center of our
HAWKI field is about [Fe/H]=-0.3, very close to the mean metallicity of classical Cepheids in the LMC.
With a reasonable expected deviation of the mean metallicity of our current M33 Cepheid sample from the mean
metallicity of LMC Cepheids by $\pm$ 0.1 dex, and an assumed metallicity sensitivity of the near-infrared
PL relations of -0.1 $\pm$ 0.1 mag/dex (Storm et al. 2011b), which is in line with the recent result
of Freedman \& Madore (2011), the expected effect of a metallicity difference
between our M33 Cepheid sample and the LMC Cepheids defining the fiducial PL relations on the distance
modulus of M33 is in the order of just $\pm$ 0.01-0.02 mag. Metallicity effects on our present determination
of the M33 true distance modulus seem therefore negligible.

\subsection{Crowding and Blending Effects}
At a distance of 839 kpc, corresponding to our present result for M33, the possible over-brightening
of some of the observed Cepheids due to relatively bright companion stars unresolved in the photometry
is clearly a potential problem,
particularly in ground-based observations. A good way to estimate the expected effect is to use the 
investigation of Bresolin et al. (2005) of the effect of blending on Cepheid magnitudes in NGC 300, a galaxy
very similar to M33 in mass, stellar density and inclination with a distance being a factor of 2.2 larger
than M33 (Gieren et al. 2005b). In NGC 300, the effect of blending of Cepheids on the distance modulus was
determined to be 0.04 mag if ground-based BVI images obtained at a 2.2 m telescope (the ESO 2.2 m telescope 
and Wide Field Imager) are used to obtain the Cepheid photometry. In the case of the present near-infrared
photometry, which strongly dominates our derived distance modulus of M33, blending should be less a problem
given the larger spatial resolution obtained with the VLT/HAWKI telescope and imager. On the other hand,
blending with the numerous bright, red AGB stars might become a more worrisome problem, but probably is
only becoming a dominating source of error at mid-infrared wavelengths (Freedman et al. 2009).
In the Bresolin et al. 2005 paper, Cepheids in outer fields in NGC 300 showing a 
similar stellar density as our present VLT/HAWKI field were used for the determination of the impact
of blending on the distance modulus.

Scaling the Bresolin et al. result for NGC 300 to the shorter distance of M33, we would expect 
a $\sim$0.02 mag
effect (systematic in the sense to make Cepheids brighter, thus underestimating the true distance of
M33). This is in line with the results of the recent study of Chavez et al. (2012) who compared VI photometry for 149
Cepheids in M33 obtained with the WIYN 3.5-m telescope to archival HST data, and reached conclusions
about the effect of blending on the Cepheid magnitudes which are very similar, and consistent with those of Bresolin
et al. (2005).  

The effect of blending through companion stars unresolved in the photometry may be expected to affect
shorter-period and fainter Cepheids more strongly than the brighter, long-period Cepheids, which
was a principal reason to restrict our sample and analysis to the brighter Cepheids available
for study in our selected field. As metallicity effects, we conclude that blending of the Cepheids
in our sample produces a minor systematic effect on the distance modulus of M33 in the order
of 1-2\%.

\section{Summary and Conclusions}
Motivated by the amazing range of distance values reported in the past decade for our Local Group spiral
neigbor galaxy M33 which is a prime anchor point for the calibration of the extragalactic distance scale,
we have performed a new distance determination for M33 based on random-phase near-infrared
photometry of a sizeable number of long-period classical Cepheid variables. The data were obtained with the
ESO VLT telescope and HAWKI near-infrared imager. Our near-infrared photometry, in conjunction with
the previous optical V,I photometry of Macri et al. (2001) of the variables has allowed us to accurately 
determine the mean reddening affecting the Cepheid sample, and the true distance modulus of the galaxy.
The total uncertainty of of the true distance modulus is found to be $\pm$0.07 mag, or 4\%. Our distance 
modulus
of 24.62 mag shows excellent agreement with the value derived by the HST Key Project (Freedman et al. 2001),
and is based on a near-identical (0.19 vs. 0.20) value of the mean reddening. There is also excellent agreement
with the more recent determination from Cepheids by Scowcroft et al. (2009) when their assumed LMC distance 
of 18.40 is corrected to 18.50, the value assumed in our present and all other Cepheid-based distance 
determinations to M33 in Table 1, and to the recent determination of 24.64 mag derived from ground-based VI photometry
of more than 500 Cepheids by Pellerin \& Macri (2011; their Table 2, cleaned sample).

We show that the effects of reddening and the assumed reddening law,
and the effect of metallicity on our derived Cepheid distance are very small; 
the former as a result of using accurate near-infrared photometry, particularly K-band photometry,
 with its relative insensitivity to reddening
in conjunction with optical photometry to extend the wavelength baseline, the latter as a result of
choosing a field containing Cepheids of similar metallicity as the LMC Cepheids which provide the fiducial
period-luminosity relations in VIJK we are using. We briefly discuss the expected effect of possible blending
of the Cepheids in our M33 sample on our distance result and argue that by comparison with a former
study of NGC 300, and recent work on M33 itself, the effect on the present M33 distance should be in the order 
of just 1-2\%. The main contributor to
the total uncertainty of the present distance determination to M33 is the uncertainty on the photometric
zero points ($\pm$ 2.5\%).

Finally, we note that our derived distance value for M33 is tied to an assumed LMC true distance modulus of 
18.50. While some recent work has suggested that this value may need a revision towards a shorter distance 
in the range
18.40-18.48 mag (Fouqu{\'e} et al. 2007; Freedman \& Madore 2010; Storm et al. 2011a; Ripepi et al. 2012;
Walker 2012), the very recent work of our group on eight eclipsing binary systems in the LMC composed of pairs of
red giants has led to a very accurate (2.2\%) distance determination of 18.497 mag for the LMC barycenter 
(Pietrzynski et al. 2013), in excellent agreement with the value we have adopted in our Araucaria Project in
all previous distance determinations to galaxies in the Local Group and beyond reported in this project.

\acknowledgments
We gratefully acknowledge financial support for this work from the BASAL Centro
de Astrofisica y Tecnologias Afines (CATA) PFB-06/2007. Support from the
Ideas Plus program of Polish Ministry of Science and Higher
Education, and the TEAM subsidy of the
Foundation for Polish Science (FNP) is also acknowledged. AG acknowledges support
from FONDECYT grant No. 3130361. We greatly appreciate the expert support
of the ESO staff at Paranal Observatory where the data for this project were obtained.

\begin{figure}[p] 
\vspace*{18cm}
\includegraphics{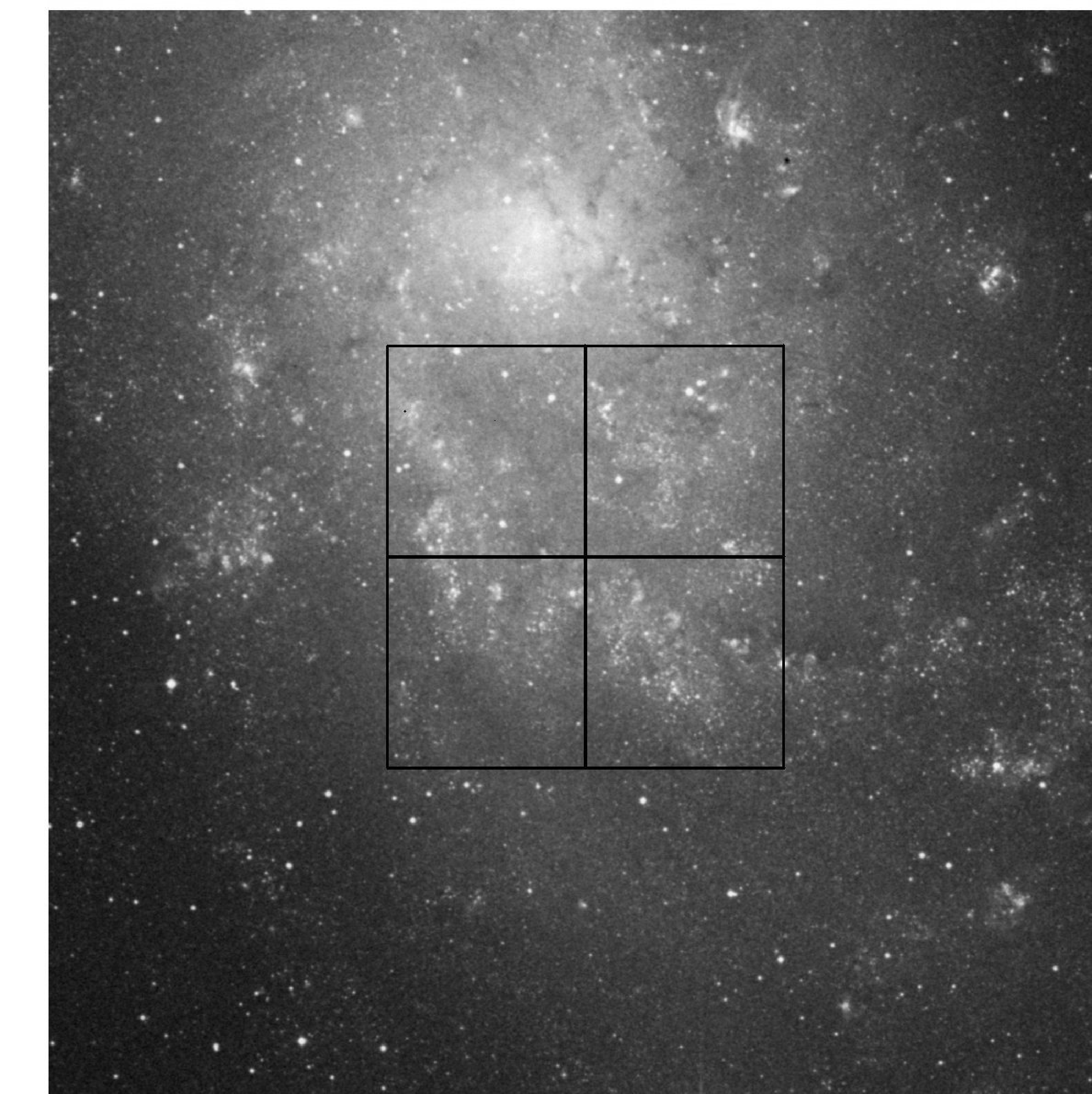} 
\caption{The location of the observed VLT/HAWKI field in M33. North is up, East to the left.
}
\end{figure}  

\begin{figure}[htb]
\vspace*{16cm}
\includegraphics{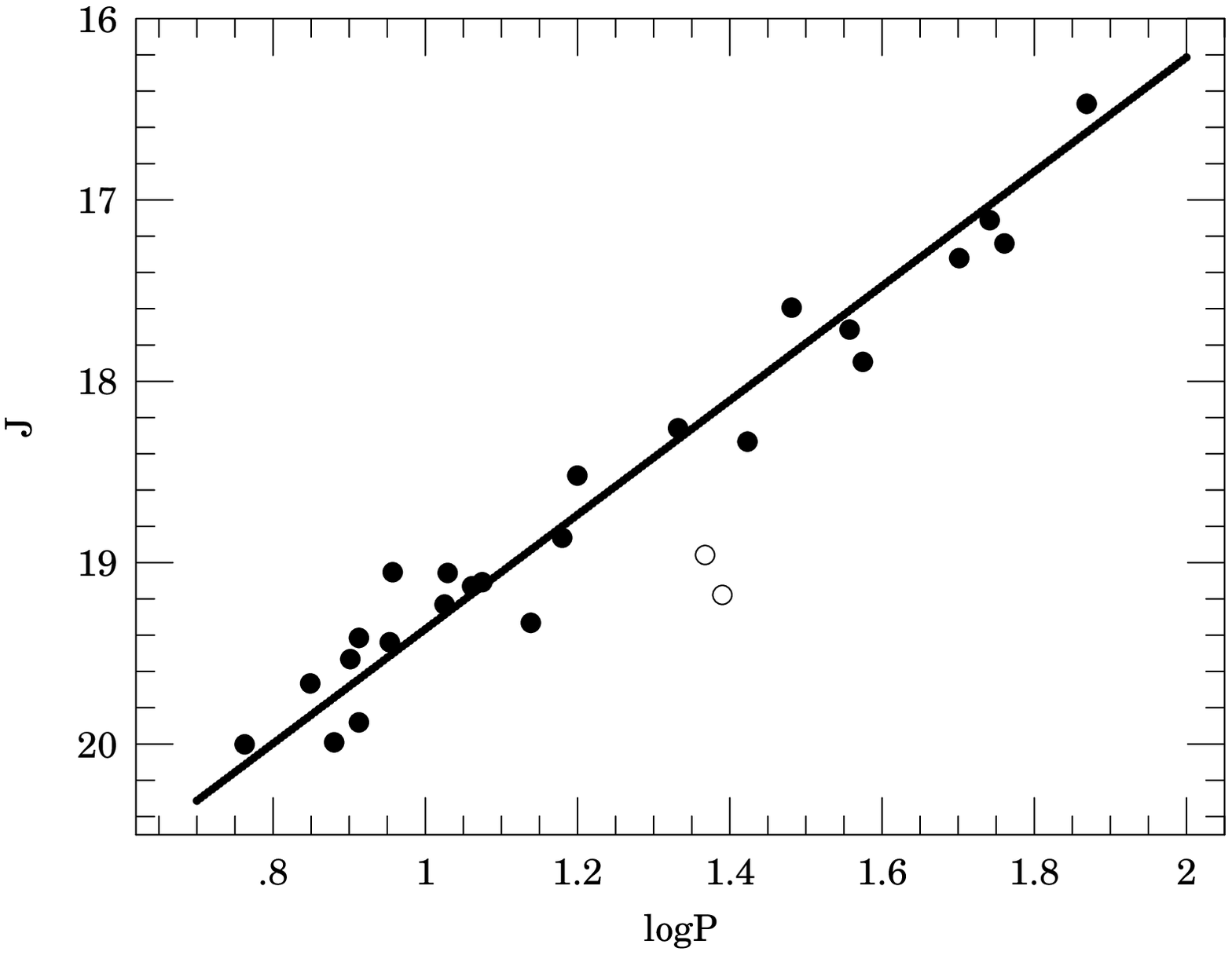}
\includegraphics{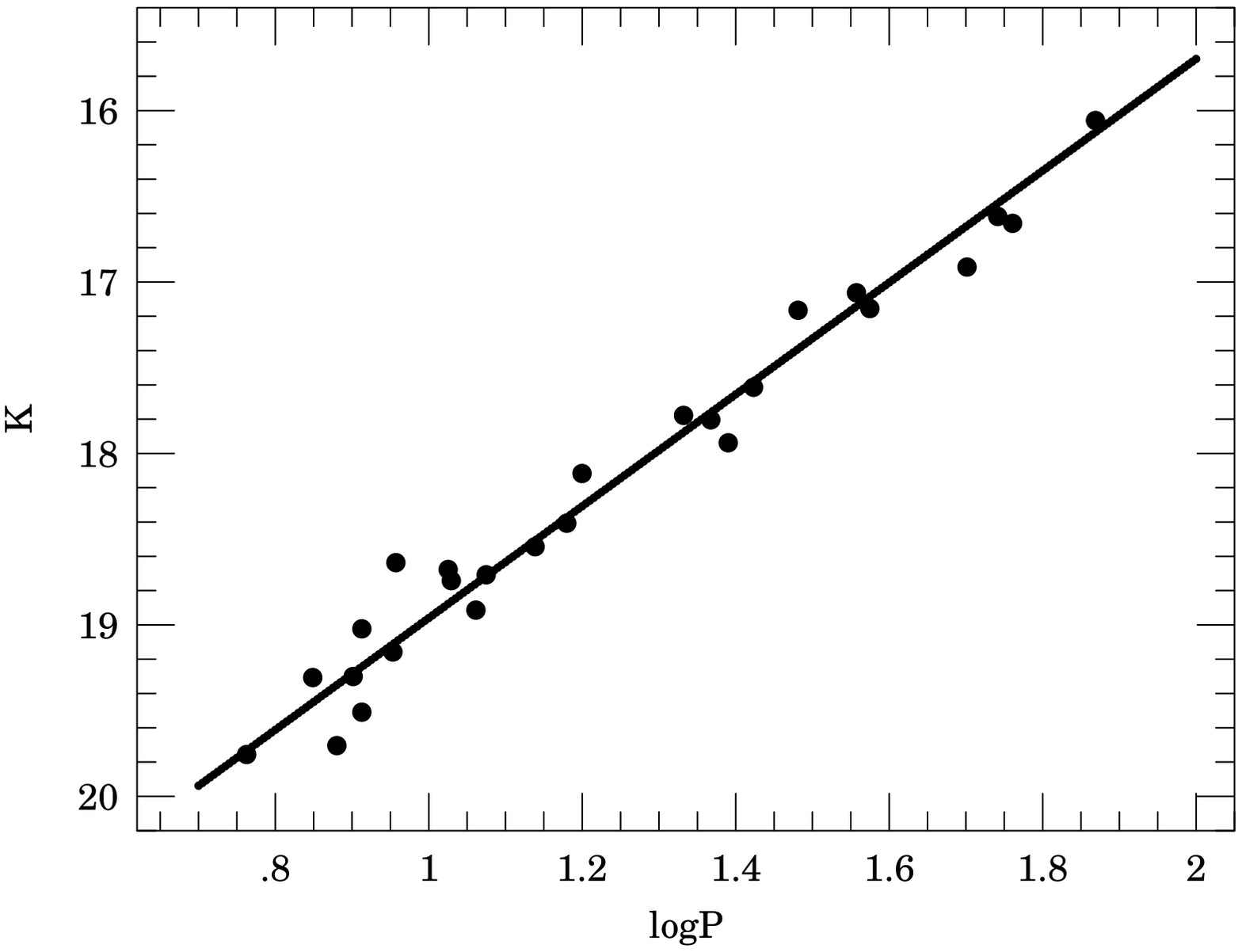}
\caption{The J-band (lower panel) and K-band (upper panel) PL diagrams in M33 obtained
 from the 26 classical Cepheids identified in our HAWK-I field. The K mean magnitudes were
 obtained from averaging 2-5 random-phase observations, whereas the J magnitudes are from one single 
 random-phase observation (see Table 2). The least-squares fits to a line assume the slopes taken from 
  LMC Cepheids, and fit the M 33 data very well. In the J-band PL
 diagram, the two Cepheids marked by open circles were not used in the analysis given their large
 deviation from the mean relation defined by the other stars. Potential reasons for the excessive faintness
 of these two objects include that the observations were made when these Cepheids were close to minimum
 brightness, and their reddenings being larger than the mean reddening derived from Fig. 3. Their inclusion
 in the analysis would not cause any significant change in the distance result for M33.
}
\end{figure}

\begin{figure}[p]
\includegraphics{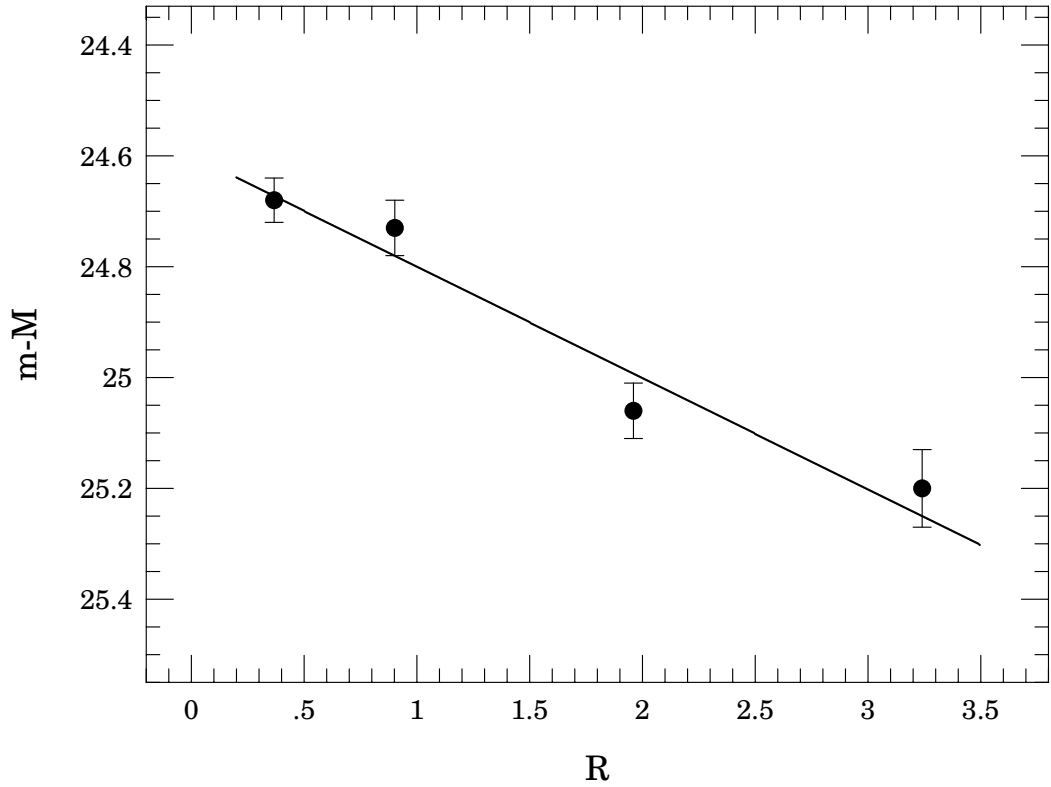}
\vspace{10cm}
\caption{Apparent distance moduli to M33 as derived in the VIJK photometric bands,
plotted against the ratio of total to selective extinction as adopted from
the Schlegel et al. reddening law. The intersection and
slope of the best-fitting line give the true distance modulus and the average total
reddening, respectively. }
\end{figure}

\begin{figure}[p]
\vspace*{18cm}
\includegraphics{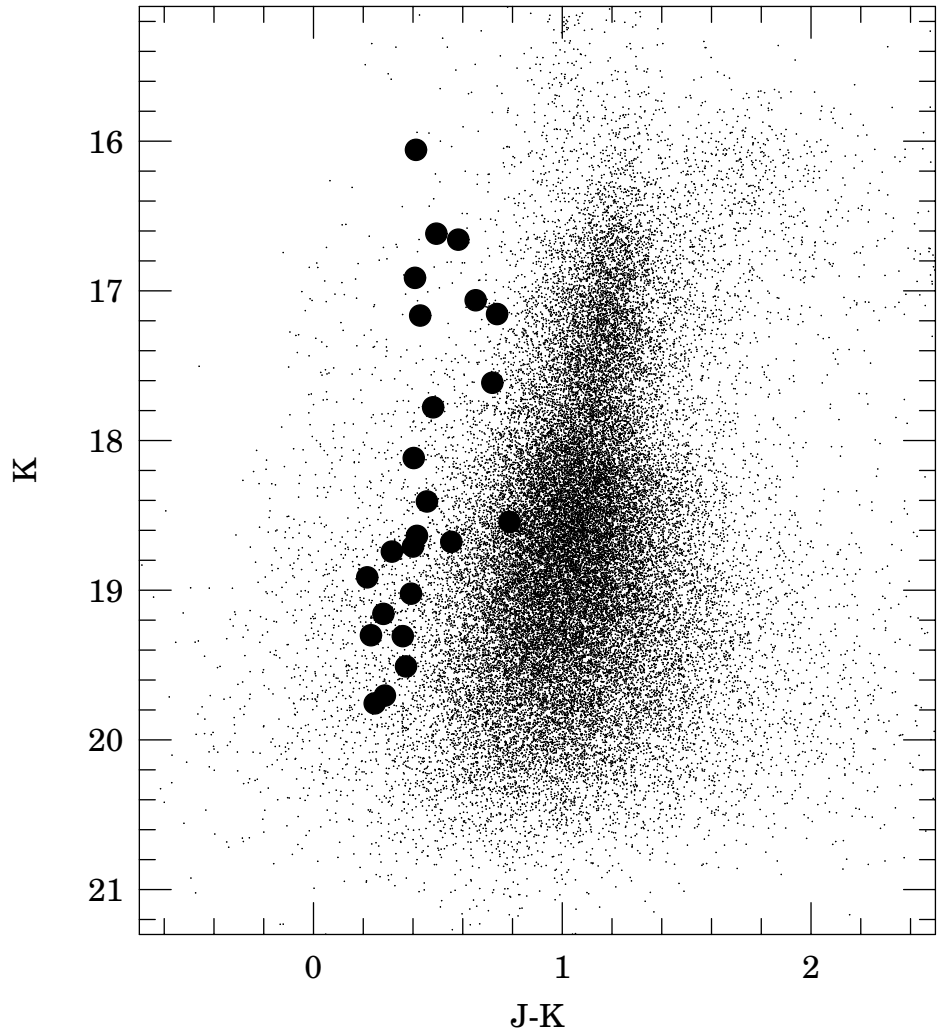}
\caption{Location of the Cepheids observed in our study on the K, J-K color-magnitude diagram obtained
from our photometry. The Cepheids are located in the expected region on this diagram, confirming
their nature as classical Cepheids. 
}
\end{figure}

\clearpage
\begin{deluxetable}{c c c}
\tablecaption{Distance Determinations to M33}
\tablehead{ \colhead{Method} & \colhead{True Distance Modulus} & \colhead{Reference}
 }
\startdata
Cepheids, BVRI (ground) &       24.64 $\pm$ 0.09 &                      Freedman et al. 1991 \\
Cepheids, VI (ground)   &     24.56 $\pm$ 0.10   &                  Freedman et al. 2001 \\
Cepheids, I, HST        &     24.52 $\pm$ 0.14 (r) $\pm$ 0.13 (s) &     Lee et al. 2002 \\
Cepheids, BVI  (ground) &     24.53 $\pm$ 0.11                  &   Scowcroft et al. 2009 \\
Cepheids, NIR (H) (ground)&    24.3 $\pm$ 0.20                  &     Madore et al. 1985  \\
Horizontal Branch         &   24.84 $\pm$ 0.16                  &    Sarajedini et al. 2000\\
TRGB in I-band  (HST)     &    24.81 $\pm$ 0.04 (r) $\pm$ 0.13 (s) &     Kim et al. 2002\\
Red Clump in I  (HST)     &    24.80 $\pm$ 0.04 (r) $\pm$ 0.05 (s) &    Kim et al. 2002\\
PNLF                      &    24.86 $\pm$ 0.09                  &   Ciardullo et al. 2004\\
H2O masers,  VLBA         &    24.32 $\pm$ 0.45                  &   Brunthaler et al. 2005\\
O-Type Eclipsing Binary   &    24.92 $\pm$ 0.12                  &   Bonanos et al. 2006\\
RR Lyrae  (HST)           &    24.67 $\pm$ 0.07                  &   Sarajedini et al. 2006\\
TRGB in I-band  (HST)     &    24.71 $\pm$ 0.04                  &   Rizzi et al. 2007\\
TRGB in I-band  (HST)     &    24.84 $\pm$ 0.10                  &   U et al. 2009\\
Blue Supergiant FGLR      &    24.93 $\pm$ 0.11                  &   U et al. 2009\\
Cepheids, VI (ground)     &    24.64 $\pm$ 0.03 (r)              &   Pellerin \& Macri 2011\\
Cepheids, NIR (JK) (ground)&   24.62 $\pm$ 0.03 (r) $\pm$ 0.06 (s)                 &   This paper\\
\enddata
\end{deluxetable}

\clearpage
\begin{deluxetable}{c c c c c}
\tablewidth{0pc}
\tablecaption{Journal of the Individual J and K band Observations of M33
Cepheids}
\tablehead{ \colhead{ID} & \colhead{HJD} & \colhead{filter} & \colhead{mag} & \colhead{$\sigma$} \\
 }
\startdata
D33J013336.4+303437.8 & 54805.039912 & J &   19.21 &    0.03 \\ 
D33J013336.4+303437.8 & 54806.032966 & K &   18.42 &    0.01 \\ 
D33J013336.4+303437.8 & 54805.039912 & K &   18.46 &    0.01 \\ 
D33J013336.4+303437.8 & 54805.079058 & K &   18.43 &    0.01 \\ 
D33J013336.4+303437.8 & 55169.039569 & K &   18.45 &    0.02 \\ 
D33J013336.4+303437.8 & 55136.126545 & K &   18.41 &    0.04 \\ 
D33J013339.8+303412.2 & 54805.039912 & J &   19.21 &    0.03 \\ 
D33J013339.8+303412.2 & 54806.032966 & K &   18.22 &    0.01 \\ 
D33J013339.8+303412.2 & 54805.039912 & K &   18.24 &    0.01 \\ 
D33J013339.8+303412.2 & 54805.079058 & K &   18.33 &    0.01 \\ 
D33J013339.8+303412.2 & 55169.039569 & K &   18.27 &    0.02 \\ 
D33J013339.8+303412.2 & 55136.126545 & K &   18.43 &    0.05 \\ 
D33J013335.6+303649.2 & 54805.039912 & J &   19.06 &    0.03 \\ 
D33J013335.6+303649.2 & 54806.032966 & K &   18.83 &    0.01 \\ 
D33J013335.6+303649.2 & 54805.039912 & K &   18.83 &    0.01 \\ 
D33J013335.6+303649.2 & 54805.079058 & K &   18.71 &    0.01 \\ 
D33J013335.6+303649.2 & 55169.039569 & K &   18.72 &    0.03 \\ 
D33J013335.6+303649.2 & 55136.126545 & K &   18.62 &    0.05 \\ 
D33J013331.6+303704.5 & 54805.039912 & J &   19.33 &    0.03 \\ 
D33J013331.6+303704.5 & 54806.032966 & K &   18.69 &    0.01 \\ 
D33J013331.6+303704.5 & 54805.039912 & K &   18.60 &    0.01 \\ 
D33J013331.6+303704.5 & 54805.079058 & K &   18.51 &    0.01 \\ 
D33J013331.6+303704.5 & 55169.039569 & K &   18.48 &    0.03 \\ 
D33J013331.6+303704.5 & 55136.126545 & K &   18.44 &    0.05 \\ 
D33J013338.4+303602.5 & 54805.039912 & J &   18.86 &    0.02 \\ 
D33J013338.4+303602.5 & 54806.032966 & K &   18.28 &    0.01 \\ 
D33J013338.4+303602.5 & 54805.039912 & K &   18.43 &    0.01 \\ 
D33J013338.4+303602.5 & 54805.079058 & K &   18.48 &    0.01 \\ 
D33J013338.4+303602.5 & 55169.039569 & K &   18.30 &    0.02 \\ 
D33J013338.4+303602.5 & 55136.126545 & K &   18.59 &    0.05 \\ 
D33J013338.4+303602.5 & 55128.189734 & K &   18.36 &    0.06 \\ 
D33J013334.4+303530.2 & 54805.039912 & J &   18.52 &    0.02 \\ 
D33J013334.4+303530.2 & 54806.032966 & K &   18.14 &    0.01 \\ 
\enddata
\end{deluxetable}

\setcounter{table}{1}
\clearpage
\begin{deluxetable}{c c c c c}
\tablewidth{0pc}
\tablecaption{Journal of the Individual J and K band Observations of M33
Cepheids}
\tablehead{ \colhead{ID} & \colhead{HJD} & \colhead{filter} & \colhead{mag} & \colhead{$\sigma$} \\
 }
\startdata
D33J013334.4+303530.2 & 54805.039912 & K &   18.09 &    0.01 \\ 
D33J013331.5+303351.2 & 54805.039912 & J &   16.25 &    0.01 \\ 
D33J013331.5+303351.2 & 54806.032966 & K &   15.27 &    0.01 \\ 
D33J013331.5+303351.2 & 54805.039912 & K &   15.27 &    0.01 \\ 
D33J013331.5+303351.2 & 54805.079058 & K &   15.26 &    0.01 \\ 
D33J013331.5+303351.2 & 55169.039569 & K &   15.33 &    0.01 \\ 
D33J013331.5+303351.2 & 55136.126545 & K &   15.41 &    0.01 \\ 
D33J013332.9+303548.4 & 54805.039912 & J &   17.59 &    0.02 \\ 
D33J013332.9+303548.4 & 54806.032966 & K &   17.38 &    0.01 \\ 
D33J013332.9+303548.4 & 54805.039912 & K &   17.09 &    0.01 \\ 
D33J013332.9+303548.4 & 54805.079058 & K &   17.09 &    0.01 \\ 
D33J013332.9+303548.4 & 55169.039569 & K &   17.13 &    0.03 \\ 
D33J013332.9+303548.4 & 55136.126545 & K &   17.14 &    0.03 \\ 
D33J013341.2+303550.0 & 54805.039912 & J &   17.72 &    0.01 \\ 
D33J013341.2+303550.0 & 54806.032966 & K &   17.01 &    0.01 \\ 
D33J013341.2+303550.0 & 54805.039912 & K &   17.00 &    0.01 \\ 
D33J013341.2+303550.0 & 54805.079058 & K &   17.04 &    0.01 \\ 
D33J013341.2+303550.0 & 55169.039569 & K &   17.14 &    0.01 \\ 
D33J013341.2+303550.0 & 55136.126545 & K &   17.27 &    0.02 \\ 
D33J013341.2+303550.0 & 55128.189734 & K &   16.93 &    0.03 \\ 
D33J013341.6+303609.2 & 54805.039912 & J &   17.37 &    0.01 \\ 
D33J013341.6+303609.2 & 54806.032966 & K &   16.40 &    0.01 \\ 
D33J013341.6+303609.2 & 54805.039912 & K &   16.36 &    0.01 \\ 
D33J013341.6+303609.2 & 54805.079058 & K &   16.37 &    0.01 \\ 
D33J013341.6+303609.2 & 55169.039569 & K &   16.43 &    0.01 \\ 
D33J013341.6+303609.2 & 55136.126545 & K &   16.40 &    0.01 \\ 
D33J013341.6+303609.2 & 55128.189734 & K &   16.40 &    0.02 \\ 
D33J013350.9+303336.1 & 54805.039912 & J &   17.89 &    0.01 \\ 
D33J013350.9+303336.1 & 54806.032966 & K &   17.15 &    0.01 \\ 
D33J013350.9+303336.1 & 54805.039912 & K &   17.11 &    0.01 \\ 
\enddata
\end{deluxetable}

\setcounter{table}{1}
\clearpage
\begin{deluxetable}{c c c c c}
\tablewidth{0pc}
\tablecaption{Journal of the Individual J and K band Observations of M33
Cepheids}
\tablehead{ \colhead{ID} & \colhead{HJD} & \colhead{filter} & \colhead{mag} & \colhead{$\sigma$} \\
 }
\startdata
D33J013350.9+303336.1 & 54805.079058 & K &   17.28 &    0.01 \\ 
D33J013350.9+303336.1 & 55169.039569 & K &   17.09 &    0.01 \\ 
D33J013359.4+303226.7 & 54805.039912 & J &   17.32 &    0.01 \\ 
D33J013359.4+303226.7 & 54806.032966 & K &   16.81 &    0.01 \\ 
D33J013359.4+303226.7 & 54805.039912 & K &   16.81 &    0.01 \\ 
D33J013359.4+303226.7 & 54805.079058 & K &   16.81 &    0.01 \\ 
D33J013359.4+303226.7 & 55169.039569 & K &   16.89 &    0.01 \\ 
D33J013359.4+303226.7 & 55136.126545 & K &   17.06 &    0.02 \\ 
D33J013359.4+303226.7 & 55128.189734 & K &   17.10 &    0.03 \\ 
D33J013348.8+303045.0 & 54805.039912 & J &   19.41 &    0.03 \\ 
D33J013348.8+303045.0 & 54806.032966 & K &   19.12 &    0.01 \\ 
D33J013348.8+303045.0 & 54805.039912 & K &   18.97 &    0.01 \\ 
D33J013348.8+303045.0 & 54805.079058 & K &   18.92 &    0.01 \\ 
D33J013348.8+303045.0 & 55169.039569 & K &   18.98 &    0.03 \\ 
D33J013348.8+303045.0 & 55136.126545 & K &   19.13 &    0.07 \\ 
D33J013343.9+303245.1 & 54805.039912 & J &   16.47 &    0.00 \\ 
D33J013343.9+303245.1 & 54806.032966 & K &   16.00 &    0.01 \\ 
D33J013343.9+303245.1 & 54805.039912 & K &   16.00 &    0.01 \\ 
D33J013343.9+303245.1 & 54805.079058 & K &   16.00 &    0.01 \\ 
D33J013343.9+303245.1 & 55169.039569 & K &   15.99 &    0.00 \\ 
D33J013343.9+303245.1 & 55136.126545 & K &   16.19 &    0.01 \\ 
D33J013343.9+303245.1 & 55134.139710 & K &   16.17 &    0.01 \\ 
D33J013332.2+303001.9 & 54805.039912 & J &   20.00 &    0.03 \\ 
D33J013332.2+303001.9 & 54806.032966 & K &   19.88 &    0.06 \\ 
D33J013332.2+303001.9 & 55169.039569 & K &   19.63 &    0.04 \\ 
D33J013336.5+303053.2 & 54805.039912 & J &   19.99 &    0.03 \\ 
D33J013336.5+303053.2 & 54806.032966 & K &   19.47 &    0.05 \\ 
D33J013336.5+303053.2 & 54805.039912 & K &   20.01 &    0.01 \\ 
D33J013336.5+303053.2 & 54805.079058 & K &   20.06 &    0.01 \\ 
D33J013336.5+303053.2 & 55169.039569 & K &   19.67 &    0.04 \\ 
D33J013336.5+303053.2 & 55136.126545 & K &   19.61 &    0.09 \\ 
D33J013336.5+303053.2 & 55134.139710 & K &   19.41 &    0.09 \\ 
D33J013332.4+303143.3 & 54805.039912 & J &   19.53 &    0.03 \\ 
\enddata
\end{deluxetable}

\setcounter{table}{1}
\clearpage
\begin{deluxetable}{c c c c c}
\tablewidth{0pc}
\tablecaption{Journal of the Individual J and K band Observations of M33
Cepheids}
\tablehead{ \colhead{ID} & \colhead{HJD} & \colhead{filter} & \colhead{mag} & \colhead{$\sigma$} \\
 }
\startdata
D33J013332.4+303143.3 & 54806.032966 & K &   19.27 &    0.05 \\ 
D33J013332.4+303143.3 & 54805.039912 & K &   19.25 &    0.01 \\ 
D33J013332.4+303143.3 & 54805.079058 & K &   19.35 &    0.01 \\ 
D33J013332.4+303143.3 & 55169.039569 & K &   19.25 &    0.03 \\ 
D33J013332.4+303143.3 & 55134.139710 & K &   19.38 &    0.10 \\ 
D33J013336.3+303243.7 & 54805.039912 & J &   19.88 &    0.04 \\ 
D33J013336.3+303243.7 & 54806.032966 & K &   19.43 &    0.05 \\ 
D33J013336.3+303243.7 & 54805.039912 & K &   19.52 &    0.01 \\ 
D33J013336.3+303243.7 & 54805.079058 & K &   19.53 &    0.01 \\ 
D33J013336.3+303243.7 & 55169.039569 & K &   19.27 &    0.03 \\ 
D33J013336.3+303243.7 & 55136.126545 & K &   19.80 &    0.13 \\ 
D33J013337.5+303305.1 & 54805.039912 & J &   19.44 &    0.03 \\ 
D33J013337.5+303305.1 & 54806.032966 & K &   19.07 &    0.04 \\ 
D33J013337.5+303305.1 & 54805.039912 & K &   19.25 &    0.01 \\ 
D33J013337.5+303305.1 & 54805.079058 & K &   19.15 &    0.01 \\ 
D33J013341.9+302951.8 & 54805.039912 & J &   19.23 &    0.03 \\ 
D33J013341.9+302951.8 & 54806.032966 & K &   18.83 &    0.05 \\ 
D33J013341.9+302951.8 & 54805.039912 & K &   18.73 &    0.01 \\ 
D33J013341.9+302951.8 & 54805.079058 & K &   18.47 &    0.01 \\ 
D33J013335.5+303330.2 & 54805.039912 & J &   19.13 &    0.03 \\ 
D33J013335.5+303330.2 & 54806.032966 & K &   18.81 &    0.04 \\ 
D33J013335.5+303330.2 & 54805.039912 & K &   18.92 &    0.01 \\ 
D33J013335.5+303330.2 & 54805.079058 & K &   18.92 &    0.01 \\ 
D33J013335.5+303330.2 & 55169.039569 & K &   19.00 &    0.03 \\ 
D33J013337.7+303218.9 & 54805.039912 & J &   19.11 &    0.02 \\ 
D33J013337.7+303218.9 & 54806.032966 & K &   18.60 &    0.03 \\ 
D33J013337.7+303218.9 & 54805.039912 & K &   18.73 &    0.01 \\ 
D33J013337.7+303218.9 & 54805.079058 & K &   18.84 &    0.01 \\ 
D33J013337.7+303218.9 & 55169.039569 & K &   18.63 &    0.02 \\ 
D33J013337.7+303218.9 & 55136.126545 & K &   18.70 &    0.05 \\ 
D33J013337.7+303218.9 & 55134.139710 & K &   18.74 &    0.07 \\ 
D33J013335.0+303336.5 & 54805.039912 & J &   18.82 &    0.02 \\ 
D33J013335.0+303336.5 & 54806.032966 & K &   16.90 &    0.02 \\ 
D33J013335.0+303336.5 & 54805.079058 & K &   16.99 &    0.01 \\ 
\enddata
\end{deluxetable}

\setcounter{table}{1} 
\clearpage
\begin{deluxetable}{c c c c c}
\tablewidth{0pc}
\tablecaption{Journal of the Individual J and K band Observations of M33
Cepheids}
\tablehead{ \colhead{ID} & \colhead{HJD} & \colhead{filter} & \colhead{mag} & \colhead{$\sigma$} \\
 }
\startdata
D33J013335.0+303336.5 & 55169.039569 & K &   17.00 &    0.01 \\ 
D33J013331.8+302957.8 & 54805.039912 & J &   18.26 &    0.03 \\ 
D33J013331.8+302957.8 & 54806.032966 & K &   17.88 &    0.05 \\ 
D33J013331.8+302957.8 & 54805.039912 & K &   17.74 &    0.01 \\ 
D33J013331.8+302957.8 & 54805.079058 & K &   17.77 &    0.01 \\ 
D33J013331.8+302957.8 & 55169.039569 & K &   17.69 &    0.03 \\ 
D33J013331.8+302957.8 & 55136.126545 & K &   17.63 &    0.05 \\ 
D33J013331.8+302957.8 & 55134.139710 & K &   17.95 &    0.05 \\ 
D33J013331.1+303143.0 & 54805.039912 & J &   17.11 &    0.00 \\ 
D33J013331.1+303143.0 & 54806.032966 & K &   16.58 &    0.01 \\ 
D33J013331.1+303143.0 & 54805.039912 & K &   16.61 &    0.02 \\ 
D33J013331.1+303143.0 & 54805.079058 & K &   16.61 &    0.01 \\ 
D33J013331.1+303143.0 & 55169.039569 & K &   16.76 &    0.01 \\ 
D33J013331.1+303143.0 & 55136.126545 & K &   16.58 &    0.02 \\ 
D33J013331.1+303143.0 & 55134.139710 & K &   16.57 &    0.01 \\ 
D33J013337.5+303138.5 & 54805.039912 & J &   17.24 &    0.01 \\ 
D33J013337.5+303138.5 & 54806.032966 & K &   16.84 &    0.01 \\ 
D33J013337.5+303138.5 & 54805.039912 & K &   16.77 &    0.01 \\ 
D33J013337.5+303138.5 & 54805.079058 & K &   16.77 &    0.01 \\ 
D33J013337.5+303138.5 & 55169.039569 & K &   16.51 &    0.01 \\ 
D33J013337.5+303138.5 & 55136.126545 & K &   16.53 &    0.02 \\ 
D33J013337.5+303138.5 & 55134.139710 & K &   16.53 &    0.02 \\ 
D33J013350.6+303445.8 & 54805.039912 & J &   19.63 &    0.03 \\ 
D33J013350.6+303445.8 & 54806.032966 & K &   18.57 &    0.03 \\ 
D33J013350.6+303445.8 & 54805.039912 & K &   18.62 &    0.01 \\ 
D33J013350.6+303445.8 & 54805.079058 & K &   18.63 &    0.01 \\ 
D33J013350.6+303445.8 & 55169.039569 & K &   18.64 &    0.02 \\ 
D33J013350.6+303445.8 & 55136.126545 & K &   18.76 &    0.06 \\ 
D33J013402.5+303628.0 & 54805.039912 & J &   18.33 &    0.03 \\ 
D33J013402.5+303628.0 & 54806.032966 & K &   17.21 &    0.01 \\ 
D33J013402.5+303628.0 & 54805.039912 & K &   17.12 &    0.01 \\ 
\enddata
\end{deluxetable}

\setcounter{table}{1}
\clearpage
\begin{deluxetable}{c c c c c}
\tablewidth{0pc}
\tablecaption{Journal of the Individual J and K band Observations of M33
Cepheids}
\tablehead{ \colhead{ID} & \colhead{HJD} & \colhead{filter} & \colhead{mag} & \colhead{$\sigma$} \\
 }
\startdata
D33J013402.5+303628.0 & 54805.079058 & K &   17.13 &    0.01 \\ 
D33J013402.5+303628.0 & 55169.039569 & K &   17.18 &    0.03 \\ 
D33J013402.5+303628.0 & 55136.126545 & K &   17.13 &    0.03 \\ 
D33J013350.7+303544.2 & 54805.039912 & J &   18.96 &    0.02 \\ 
D33J013350.7+303544.2 & 54806.032966 & K &   17.65 &    0.03 \\ 
D33J013350.7+303544.2 & 54805.039912 & K &   17.81 &    0.01 \\ 
D33J013350.7+303544.2 & 54805.079058 & K &   17.85 &    0.01 \\ 
D33J013350.7+303544.2 & 55169.039569 & K &   17.80 &    0.01 \\ 
D33J013350.7+303544.2 & 55136.126545 & K &   17.92 &    0.03 \\ 
D33J013358.8+303719.7 & 54805.039912 & J &   19.18 &    0.03 \\ 
D33J013358.8+303719.7 & 54806.032966 & K &   17.99 &    0.03 \\ 
D33J013358.8+303719.7 & 54805.039912 & K &   17.90 &    0.01 \\ 
D33J013358.8+303719.7 & 54805.079058 & K &   17.90 &    0.01 \\ 
D33J013358.8+303719.7 & 55169.039569 & K &   17.95 &    0.03 \\ 
D33J013358.8+303719.7 & 55136.126545 & K &   17.96 &    0.04 \\ 
D33J013347.2+303536.2 & 54805.039912 & J &   18.33 &    0.02 \\ 
D33J013347.2+303536.2 & 54806.032966 & K &   17.69 &    0.02 \\ 
D33J013347.2+303536.2 & 54805.039912 & K &   17.66 &    0.01 \\ 
D33J013347.2+303536.2 & 54805.079058 & K &   17.66 &    0.01 \\ 
D33J013347.2+303536.2 & 55169.039569 & K &   17.51 &    0.01 \\ 
D33J013347.2+303536.2 & 55136.126545 & K &   17.56 &    0.03 \\ 
\enddata
\end{deluxetable}

\clearpage
\begin{deluxetable}{c c c c}
\tablewidth{0pc}
\tablecaption{Adopted Mean J and K band Magnitudes of M33
Cepheids}
\tablehead{ \colhead{ID} & \colhead{J} & \colhead{K} & \colhead{log P} \\
\colhead{} & \colhead{mag} & \colhead{mag} & \colhead{days}
 }
\startdata
D33J013332.2+303001.9 &   20.00 &   19.76 &    0.76 \\ 
D33J013332.5+303408.9 &   19.67 &   19.31 &    0.85 \\ 
D33J013336.5+303053.2 &   19.99 &   19.70 &    0.88 \\ 
D33J013332.4+303143.3 &   19.53 &   19.30 &    0.90 \\ 
D33J013336.3+303243.7 &   19.88 &   19.51 &    0.91 \\ 
D33J013348.8+303045.0 &   19.41 &   19.02 &    0.91 \\ 
D33J013337.5+303305.1 &   19.44 &   19.16 &    0.95 \\ 
D33J013336.8+303434.4 &   19.05 &   18.64 &    0.96 \\ 
D33J013341.9+302951.8 &   19.23 &   18.68 &    1.03 \\ 
D33J013335.6+303649.2 &   19.06 &   18.74 &    1.03 \\ 
D33J013335.5+303330.2 &   19.13 &   18.91 &    1.06 \\ 
D33J013337.7+303218.9 &   19.11 &   18.71 &    1.07 \\ 
D33J013331.6+303704.5 &   19.33 &   18.54 &    1.14 \\ 
D33J013338.4+303602.5 &   18.86 &   18.41 &    1.18 \\ 
D33J013334.4+303530.2 &   18.52 &   18.12 &    1.20 \\ 
D33J013331.8+302957.8 &   18.26 &   17.78 &    1.33 \\ 
D33J013350.7+303544.2 &   18.96 &   17.80 &    1.37 \\ 
D33J013358.8+303719.7 &   19.18 &   17.94 &    1.39 \\ 
D33J013347.2+303536.2 &   18.33 &   17.61 &    1.42 \\ 
D33J013332.9+303548.4 &   17.59 &   17.17 &    1.48 \\ 
D33J013341.2+303550.0 &   17.72 &   17.06 &    1.56 \\ 
D33J013350.9+303336.1 &   17.89 &   17.16 &    1.57 \\ 
D33J013359.4+303226.7 &   17.32 &   16.91 &    1.70 \\ 
D33J013331.1+303143.0 &   17.11 &   16.62 &    1.74 \\ 
D33J013337.5+303138.5 &   17.24 &   16.66 &    1.76 \\ 
D33J013343.9+303245.1 &   16.47 &   16.06 &    1.87 \\ 
\enddata
\end{deluxetable}
\clearpage
\begin{deluxetable}{cccccc}
\tablewidth{0pc}
\tablecaption{Reddened and Absorption-Corrected Distance Moduli for NGC
M33 in Optical and Near-Infrared Bands}
\tablehead{ \colhead{Band} & $V$ & $I$ & $J$ & $K$ & $E(B-V)$ }
\startdata
$m-M$                &   25.20 &  25.06 &  24.73 &  24.70 &   --  \nl
${\rm R}_{\lambda}$  &   3.24   &  1.96   &  0.902  &  0.367  &   --  \nl
$(m-M)_{0}$          &   24.58  &  24.68 &  24.56 &  24.63 &  0.19 \nl
\enddata
\end{deluxetable}

\end{document}